\documentclass[12pt]{article}

\usepackage{float}
\usepackage{amsmath}
\usepackage{epsfig}
\usepackage{graphicx}
\usepackage{subfigure}
\renewcommand{\in}{\raise -3pt\hbox{\scriptsize in}}
\newcommand{\out}{\raise -3pt\hbox{\scriptsize out}}

\newcommand{\re}{\mathrm{Re}\,}
\newcommand{\im}{\mathrm{Im}\,}

\newcommand{\ep}{\varepsilon}

\newcommand{\eqs}[1]{\begin{equation} \begin{split} #1\end{split} \end{equation} }

\newcommand{\ks}[1]{#1 \!\!\! \slash } 

\newcommand{\ga}{\gamma^5}
\newcommand{\gmu}{\gamma^\mu}

\newcommand{\tra}{{\rm Tr}}
\newcommand{\ie}{{\it i.e.}}
\newcommand{\eg}{{\it e.g.}}
\newcommand{\etal}{{\it et al.}}

\newcommand{\ce}[1]{Eq.~(\ref{#1})}

\newcommand{\cf}[1]{{Fig.~\ref{#1}}}

\newcommand{\nn}{\nonumber}

\newcommand{\beq}[1]{
\begin{equation}\label{#1}}
\newcommand{\eeq}{\end{equation}}
\newcommand{\bea}[1]{
\begin{eqnarray}\label{#1}}
\newcommand{\eea}{\end{eqnarray}}
\begin{document}

\begin{flushright}
  CPHT-RR-012.0206 \\
  hep-ph/0602195
\end{flushright}
\vspace{\baselineskip}

\begin{center}
\textbf{\LARGE Exclusive meson pair production in $\gamma^\star\gamma $
 scattering at small momentum transfer} \\

%
\vspace{1\baselineskip}
{\large
 J.P.~Lansberg $^{a,b}$,\, B.~Pire$^{a}$,\,
 L.~Szymanowski$^{b,c,d}$
}
\\
\vspace{2\baselineskip}
${}^a$\,CPHT\footnote{Unit{\'e} mixte 7644 du CNRS}, {\'E}cole
Polytechnique, 91128 Palaiseau, France \\[0.5\baselineskip]
${}^b$\,Physique Th\'eorique Fondamentale, Universit\'e de Li\`ege,\\
17 All\'ee du 6 Ao\^ut, B\^at. B5, B-4000 Li\`ege-1, Belgium\\[0.5\baselineskip]
${}^c$\,Soltan Institute for Nuclear Studies, Warsaw, Poland \\[0.5\baselineskip]
${}^d$\,LPT\footnote{Unit\'e mixte 8627 du CNRS }, Universit\'e Paris-Sud, 91405, Orsay, France

\vspace{2\baselineskip}
{\it Emails: Jean-Philippe.Lansberg@cpht.polytechnique.fr, pire@cpht.polytechnique.fr, lech@qcd.theo.phys.ulg.ac.be }

\vspace{2\baselineskip}
\textbf{Abstract}\\
\vspace{1\baselineskip}
\parbox{0.9\textwidth}
{We  study  the exclusive production of $\pi\pi$  and $\rho\pi$ 
in hard $\gamma^\star \gamma$ scattering in the forward 
kinematical region where the virtuality of one photon provides us with a hard scale in the process. 
The newly  introduced concept of Transition Distribution Amplitudes  (TDA) is used to perform a QCD 
calculation of these reactions thanks to two simple models for TDAs. 
Cross sections for  $ \rho \pi$ and $\pi \pi $  production are 
evaluated and compared to the possible background from the Bremsstrahlung process. This picture
 may be tested at intense electron-positron colliders such as CLEO and $B$ factories. The cross section 
$e\gamma~\to~e'\pi^0\pi^0$ is finally shown to provide a possible determination of the
$\pi^0$ axial form factor, $F^{\pi^0}_A$,
at small $t$, which seems not to be measurable elsewhere. }
\end{center}

\section{ Introduction}
In a recent paper \cite{PS2}, we have advocated that  factorisation theorems \cite{fact} 
for exclusive processes may be extended to the case of the reaction
$\pi^-\,\pi^+ \, \to \,\gamma^\star\,\gamma$
in the kinematical regime where the virtual photon is highly virtual (of 
the order of the energy squared of the reaction) but the momentum transfer $t$
is small. This enlarges the successful description of deep exclusive
reactions  in  terms   of  distribution  amplitudes \cite{ERBL}  and/or
generalised  parton distributions \cite{Dvcs, Dvcs2}  on the  one  side and
perturbatively   calculable  coefficient  functions   describing  hard
scattering at  the partonic  level on the  other side.  We want here to describe 
along the same lines the (crossed) reactions $\gamma_L^\star\,\gamma \to \, A B$:
   \begin{equation} 
  \gamma_L^\star\,\gamma \to \, \rho^\pm\,\pi^\mp, ~~~~ \gamma_L^\star\,\gamma \to \, \pi^\pm\,\pi^\mp, 
~~~~~~\gamma_L^\star\,\gamma \to \, \pi^0\,\pi^0, 
  \label{gagapi}
\end{equation} 
in the  near forward region and for large virtual photon invariant mass 
$Q$, which may  be studied in detail at intense electron colliders such as
 those which are mostly used as $B$ factories.

\begin{figure}[h]
\centering{\includegraphics[width=7cm]{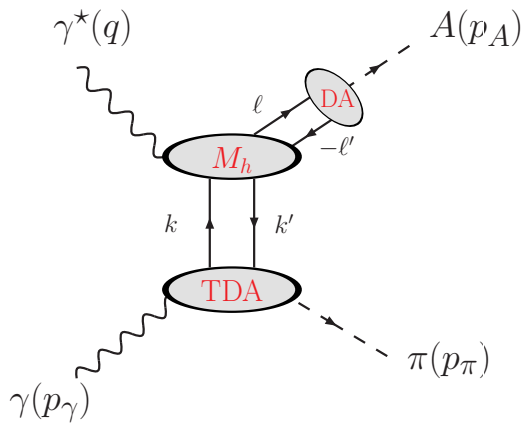}}
\caption{The factorised amplitude for $\gamma^\star \gamma \to A \pi$ at small transfer
momentum.}
\label{fig:ggstarAB}
\end{figure}

 Let us recall the main ingredients of the analysis developed in Ref. \cite{PS2, PS3}. 
 With the kinematics  described in \cf{fig:ggstarAB},  and introducing light-cone coordinates 
$v^\pm = (v^0 \pm
v^3) /\sqrt{2}$ and  transverse components $v_T = (v^1,  v^2)$ for any
four-vector $v$, we define  the $\gamma \to \pi$ transition distribution 
amplitudes (TDAs)  $T(x, \xi, t)$ as the Fourier transform of matrix elements 
$\langle     \pi(p_\pi)|\, O^\mu \,|\gamma(p_\gamma) \rangle$ where 
$O^\mu = \bar{\psi}(-\frac{z}{2})[-\frac{z}{2},\frac{z}{2}]\,\Gamma^\mu\,{\psi}(\frac{z}{2})$
 with $\Gamma^\mu = \gamma^\mu, \gamma^\mu \gamma^5,
\sigma^{\mu\,\nu}$. 
We then factorise the amplitude of the process  $  \gamma_L^\star \gamma \to A \pi$
as\footnote{The $\gamma^\star_T$ case is more difficult to analyse since the leading twist 
amplitude vanishes.}
\begin{equation}
\label{} 
{\cal  M} (Q^2, \xi, t)\propto  \int dx dz \,\Phi_{A}(z)  M_{h}(z,x,\xi) T(x, \xi, t)\;,
\end{equation}
 with a hard amplitude $M_{h}(z,x, \xi)$   
and  $\Phi_{A}(z)$   is  the   hadron  distribution
amplitude (DA).

The variable $z$ is as usual the light-cone momentum fraction carried by the quark entering the
meson $A$, $x+\xi$ (resp. $x-\xi$) is the corresponding one for the quark leaving (resp. entering) 
the TDA. The skewness variable $\xi$ describes the loss of light-cone momentum of the incident 
photon and is connected to the Bjorken 
variable $x_B$ (see section 3 for detailed kinematics).

Contrarily to the case of generalised parton distributions (GPD) 
where the forward limit is related to the conventional parton distributions 
measured in the deep inelastic scattering (DIS), there is no such 
interesting constraints for the  new TDAs.
The only constraints are sum rules obtained by taking the local limit of
the corresponding operators and possibly soft limits when the produced-meson
momentum vanishes. Lacking any non-perturbative calculations of
matrix element defining TDAs we are forced to build toy models to get 
 estimates for the cross sections, to be compared with
future experimental data.  In particular, Tiburzi~\cite{Tiburzi:2005nj} has recently 
developed a model for both these axial and vector TDAs, which we shall use in 
a simplified version to  make a comparison with the results obtained from our model. Also some other
approaches used to model GPD in the pion case~\cite{GPD_pion} could be applied 
in this context.

\section {The $ \gamma  \to\pi $ TDAs}

Let us first stress an obvious point. The $\gamma \to \pi ^\pm$ TDA involve of course only 
quark correlators, and so does the $\gamma \to \pi ^0$ TDA since the charge conjugation 
property of the $\pi^0$ forbids the leading twist gluonic TDA. Therefore, in the following, 
we only need to take into account quark correlators. 
For definiteness, let us consider the $ \gamma  \to\pi^{-} $ 
TDAs which are given by~\cite{PS2} ($P=\frac{p_{\pi^-}+p_\gamma}{2}$, 
$\Delta=p_{\pi^-}-p_\gamma$):

\begin{equation}
\begin{split}\label{eq:def_TDA}
 \int \frac{d z^-}{2\pi} e^{ix P^+ z^-} 
\langle     \pi^-(p_{\pi^-})| \bar{d}(-\frac{z}{2})\Big[-\frac{z}{2};\frac{z}{2}\Big]\gamma^\mu{u}(\frac{z}{2})|
\gamma(p_\gamma,\varepsilon) \rangle \Big|_{z^+=0,\,  z_T=0}& \\ =
\frac{1}{P^+} \frac{i\;e}{f_\pi}&\epsilon^{\mu \ep P \Delta_{\perp}} 
V^{\pi^-}(x,\xi,t),  \\
 \int \frac{d z^-}{2\pi}\, e^{ix P^+z^-}       
\langle       \pi^-(p_{\pi^-})|\bar{d}(-\frac{z}{2})\Big[-\frac{z}{2};\frac{z}{2}\Big]\gmu \ga {u}(\frac{z}{2})
|\gamma(p_\gamma,\varepsilon)    \rangle 
\Big|_{z^+=0,\,  z_T=0}  &\\  = 
\frac{1}{P^+} \frac{e}{f_\pi} &(\varepsilon\cdot\Delta) P^\mu A^{\pi^-}(x,\xi,t),
\\
 \int \frac{d z^-}{2\pi}\, e^{ix P^+z^-}
\langle       \pi^-(p_{\pi^-})|\bar{d}(-\frac{z}{2})\Big[-\frac{z}{2};\frac{z}{2}\Big]\sigma^{\mu \nu}{u}(\frac{z}{2})|
\gamma(p_\gamma,\varepsilon)    
\rangle \Big|_{z^+=0,\,  z_T=0} &\\ = \frac{e}{P^+} \epsilon^{\mu \nu \rho  
\sigma} P_\sigma \Big[ 
 \varepsilon_{\rho} 
 T^{\pi^-}_1(x,\xi,t) - \frac{1}{f_\pi} (\varepsilon \cdot \Delta) \Delta_{\perp \rho}& T^{\pi^-}_2(x,\xi,t) \Big], 
\end{split}
\end{equation}
where the first two TDAs, $V^{\pi^-}(x,\xi,t)$ and $A^{\pi^-}(x,\xi,t)$ (commonly labelled by
 $G(x,\xi,t)$ in the following) are chiral even 
and the latter ones, $T^{\pi^-}_i(x,\xi,t), i=1,2$, are chiral 
odd. In the TDA \ce{eq:def_TDA}, we include the 
Wilson line $[y;z]  \equiv   {\rm  P\   exp\,}[ig(y-z)\int_0^1\!dt\,
\,n_\mu A^\mu  (ty+(1-t)z)]$, which provides  the QCD-gauge
invariance for  non local operators  and equals unity in  a light-like
(axial) gauge. On the other hand, we do not write the 
 electromagnetic Wilson line caused by the presence of the photon, since 
we choose an electromagnetic axial gauge for the photon.

The TDAs $V^{\pi^-}(x,\xi,t)$, $A^{\pi^-}(x,\xi,t)$,  $T^{\pi^-}_1(x,\xi,t)$ and  $T^{\pi^-}_2(x,\xi,t)$
 are dimensionless quantities; $f_\pi$ is the pion decay constant ($f_\pi= 131$ MeV). These 
four leading twist TDAs are in fact linear combinations of the four independent helicity amplitudes 
for the process $q \gamma \to q \pi^-$.

\subsection{Constraints on $\gamma \to \pi$ TDAs}

   Sum rules  may  be derived  for  the  photon to meson
TDAs.  Since the local matrix elements  appear in radiative 
weak decays, we can relate them to the vector and axial 
form factors $F_V$ and $F_A$.

For $\pi^{\pm}$, for which  one has~\cite{Eidelman:2004wy}
\eqs{F_V^{\pi^{\pm}}=0.017  \pm 0.008 \hbox{  and  }
F_A^{\pi^{\pm}}=0.0116  \pm 0.0016,\label{eq:F_AV-PDG}}
 we get\footnote{In~\ce{srAV}, we use the same 
definition as in PDG~\cite{Eidelman:2004wy}, $F_G$ being a
dimensionless quantity. According to these conventions, 
the $F_G$ defined in~\cite{PS2} should be divided by
$m_\pi$.} 
\begin{equation} \int^1_0 dx \, G^{\pi^{\pm}}(x,\xi,t)= \frac{f_\pi}{m_\pi} \,F^{\pi^{\pm}}_G(t) \hspace{2cm}  G=(V,A) .
\label{srAV}
\end{equation}

To what concerns the neutral $\pi^0$, one can constrain the vector 
TDA thanks to the electromagnetic (transition) form factor $F_{\pi^0\gamma^\star \gamma}$
which is known at small $t$~\cite{Eidelman:2004wy}. Indeed, one has\footnote {This sum rule 
is slightly different from the one written in~\cite{PS2} where we had assumed a 
quark model decomposition of the $\pi^0$.}  
\begin{equation} \int^1_0 dx \left( Q^u\,V^{\pi^0}_u(x,\xi, t) 
+ Q^d\,V^{\pi^0}_d(x,\xi,t)\right)=f_\pi\,F_{\pi^0\gamma^\star \gamma} (t),
\label{srVpi0}
\end{equation}
where $Q^u=2/3$, $Q^d=-1/3$ and $V^{\pi^0}_q(x,\xi, t)$ is the TDA related to the 
operator built from the quark $q$.
Current algebra fixes the value of the right hand side at $t=0$ since 
$F_{\pi^0\gamma^\star \gamma} (t=0)= \frac{\sqrt{2}}{4\pi^2 f_\pi}$~\cite{BL81}.

These sum rules\footnote{In principle, one can also derive analogous sum rules for the chiral-odd
TDAs ($T_1$ and $T_2$ in \ce{eq:def_TDA}) but we do not see an obvious way to relate them to 
experimentally interesting observables, see the discussion at the end of section~\ref{sec:est_rho_pi}.} constrain possible parametrisations of the TDAs.
Note, in particular, the $\xi$-independence of the right hand side of the relations.

\subsection{Toy models for $\gamma \to \pi$ axial and vector TDA}

Since we want to respect the $\xi$ independence of the first moment in $x$ 
of the TDA leading to the sum rules (\ce{srAV} and \ce{srVpi0}), 
we shall use the analogy with the construction of GPDs through the double distributions~\cite{rad}.
We consider here the case of the $\gamma \to \pi^-$ TDA.

\subsubsection{$t$-independent double distributions}

In this section, we start from double distribution for $t=0$ \cite{rad}
and we omit the D-term which being isoscalar does not exist 
 for  the $\gamma \to \pi ^\pm$ TDA. A D-term may exist for the  $\gamma \to \pi ^0$ 
TDA, but we shall not include it for simplicity.

We start from the representation of the $x$ and $\xi$ dependence of the 
TDA  in the form
\beq{DD}
G^{(0)}(x,\xi)=\int_{-1}^{1} d\beta  \int_{-1+|\beta|}^{1-|\beta|}   d\alpha \; 
\delta(x-\beta -\xi \alpha)f(\beta,\alpha),
\eeq
with
\beq{f}
f(\beta,\alpha)= q(\beta)h(\beta,\alpha),
\eeq
where $q(\beta)$ is analogous to the forward quark distribution in the 
GPD case. $h(\beta,\alpha)$ is a profile function usually parametrised as
\cite{MR}
\beq{h}
h^{(b)}(\beta,\alpha)=\frac{\Gamma(2b+2)}{2^{2b+1}\Gamma^2(b+1)}
\frac{[(1-|\beta|)^2-\alpha^2]^b}{(1-|\beta|)^{2b+1}},
\eeq
where the parameter $b$ characterises the strength of the $\xi-$dependence.
As a first guess we assume that the $\beta-$dependence
of $q$ is given by a simple linear law 
\beq{q}
q(\beta) \;=\;2\;(1-\beta)\;\theta(\beta)\;.
\eeq
As usually done for GPDs, we assume a mild $\xi$ dependence as given by  $b=1$
and implement the normalisation (with $\int dx G^{(0)}(x,\xi) = 1$) and $t-$dependence of the TDA
through the axial or vector form factor:

\beq{tdep}
G(x,\xi,t) =G^{(0)}(x,\xi)\cdot \frac{f_\pi}{m_\pi}F_G(t)\;. 
\eeq
The $t-$dependence of these form factors has been studied in chiral perturbation theory \cite{bij} 
and turned out to be weak, so we shall neglect it in this model and we shall use
the measured values of \ce{eq:F_AV-PDG}.

We get
\begin{eqnarray}
G^{(0)}(x,\xi)&=&\int_0^{1} d\beta  \int_{-1+\beta}^{1-\beta}   d\alpha \; 
\frac{1}{\xi}\delta\left(\alpha-\frac{x-\beta}{\xi}\right)\frac{3}{2}
\Big[1-\frac{\alpha^2}{(1-\beta)^{2}}\Big]\nn\\
&=& \theta(x+\xi) \theta(\xi-x) \int_0^{\frac{x+\xi}{1+\xi}} d\beta\frac{3}{2\xi}
\left(1-\frac{(x-\beta)^2}{\xi^2(1-\beta)^{2}}\right)
+ \nn\\ &&(1-\theta(\xi-x)) \int_{\frac{x-\xi}{1-\xi}}^{\frac{x+\xi}{1+\xi}} d\beta\frac{3}{2\xi}
\left(1-\frac{(x-\beta)^2}{\xi^2(1-\beta)^{2}}\right)\nn\\
&=& \frac{3}{2\xi^3}\left[\theta(x+\xi) \theta(\xi-x) 
\left((x+\xi-2)(x+\xi)+2(x-1) \log\left(\frac{1-x}{1+\xi}\right)\right)
\right.
 \nn\\ &&+\left.(1-\theta(\xi-x)) 
2(x-1) \left(2\xi+ \log\left(\frac{1-\xi}{1+\xi}\right)\right)
\right].
\end{eqnarray}

We do not include any QCD evolution to our TDAs, the effects of which are supposedly 
 much less important than the uncertainty of our modelling. The resulting $\gamma \to \pi^-$ 
vector TDA is plotted on \cf{GPITDA} (a). The axial one and the ones for $\pi^0$
have the same shape with a different  normalisation. In the following, we shall refer to this
approach as Model 1.

\subsubsection{$t$-dependent double distributions}

We shall use here the initial $t$-dependent double distribution of Tiburzi~\cite{Tiburzi:2005nj}. 
Explicitly, we have:

\begin{eqnarray}
V(x,\xi,t) &=& \frac{f_\pi}{m_\pi}\int_{-1}^{1} d\beta \int_{-1 + |\beta|}^{1 - |\beta|} d\alpha \;
\delta(x - \beta - \xi \alpha) W(\beta,\alpha;t),\\
A(x,\xi,t) &=& \frac{f_\pi}{m_\pi}\int_{-1}^{1} d\beta \int_{-1 + |\beta|}^{1 - |\beta|} d\alpha \;
\delta(x - \beta - \xi \alpha) B(\beta,\alpha;t),
\end{eqnarray}
with 
\begin{eqnarray}
W(\beta,\alpha;t)&=&
\frac{m^2}{4 \pi^2} 
\Bigg\{ 
m^2 + \frac{1}{2} \beta ( \alpha + \beta - 1) m_\pi^2  - [1 - \alpha^2 - \beta ( 2 - \beta)] \frac{t}{4}
\Bigg\}^{-1}\hspace*{-.3cm},
\\
B(\beta,\alpha;t)&=&\frac{3 m^2 ( \alpha + \beta) }{4 \pi^2} 
\Bigg\{ 
m^2 + \frac{1}{2} \beta ( \alpha + \beta - 1) m_\pi^2  - [1 - \alpha^2 - \beta ( 2 - \beta)] \frac{t}{4}
\Bigg\}^{-1}\hspace*{-.3cm},
\end{eqnarray}
where $m$ is set to $0.18$ GeV~\cite{Tiburzi:2005nj}. The sum rule for $V$ is satisfied 
for $t=-0.5$ GeV$^2$. We shall keep
$t$ fixed in the following to enable comparison with Model 1. In the following, 
we shall refer  to this approach as Model 2. We see that these two models give quite
 different results and we shall use both 
in section 3 to estimate the sensitivity of the cross sections to TDA models.

\begin{figure}[h]
\centering
\mbox{\subfigure{\includegraphics[width=8.5cm]{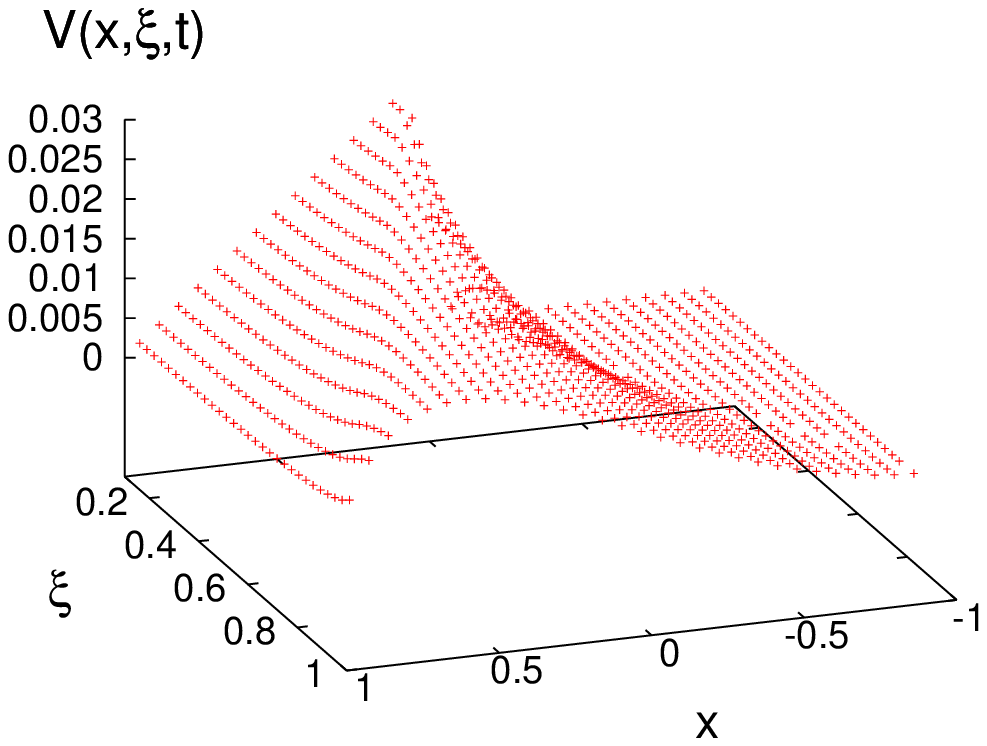}}
      \subfigure{\includegraphics[width=8.5cm]{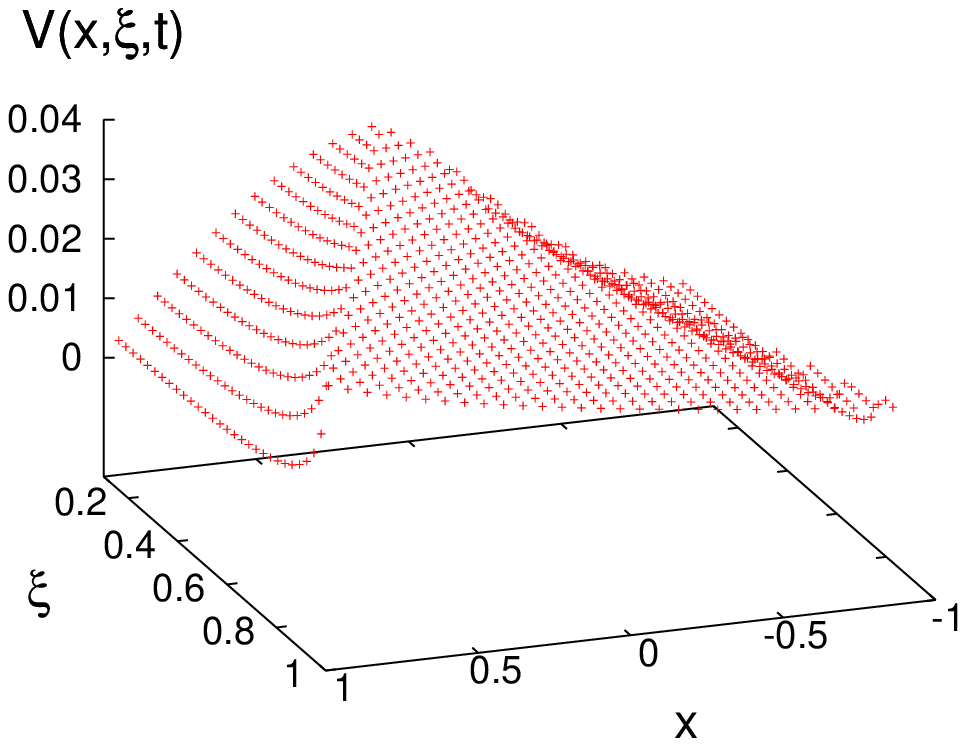}}}
\caption{The  $\gamma \to \pi^{-}$  vector transition distribution amplitude $V(x,\xi,t)$ in Model 1 
and in Model 2 (for $t=-0.5$ GeV$^2$).}
\label{GPITDA}
\end{figure}


 \section{$e \gamma \to e \rho_L \,\pi$ cross-section  estimates}\label{sec:est_rho_pi}

We first consider the $\rho_L\,\pi$ production case and more precisely when the $\rho$ flies 
in the direction of the virtual photon and the $\pi$ enters the TDA. It is 
cleaner than the $\pi^\pm \,\pi^\mp$ production case. Indeed, the reaction 
$\gamma^\star\,\gamma \to \, A B $ is accessible only in a $e\, \gamma$ 
reaction\footnote{We do not consider the complete case of $e^+\,e^-$ 
collisions which may as usual be rewritten in the equivalent
photon approximation in terms of the $e\, \gamma$ reaction and the quasi
real photon flux.}
and thus must compete with the Bremsstrahlung process 
where the meson pair $A\pi$ comes from a virtual photon emitted by the lepton line. 
In the $\rho_L \,\pi$ case, this Bremsstrahlung process is much suppressed because of
the vector and axial-vector character of the $\rho_L$ and $\pi$ leading-twist DAs; this leads to
a vanishing leading-twist contribution to the form factor $F_{\rho\pi}$ at high
transfer momentum. Moreover, in the neutral $\rho^0 \,\pi^0$  case, the TDA process 
is forbidden by $C$-conjugation.

\subsection{Kinematics}

The momenta in the process $e\gamma\to e\rho^\pm\pi^\mp$ are defined as shown on \cf{fig:kinematics}
in the CMS of the $\rho^\pm$ and $\pi^\mp$ (or equally of the $\gamma$ and $\gamma^\star$). We choose
the $x-z$ plane to be the one of the collisions $\gamma^\star\,\gamma \to \rho \,\pi$, which we call 
hadronic plane. The leptonic one,~\ie~where the $\gamma^\star$ is emitted by the electron is at 
an angle $\varphi$ of the hadronic one. We also have $P=\frac{p_{\pi^\mp}+p_\gamma}{2}$, 
$\Delta=p_{\pi^\mp}-p_\gamma$ and $q=p_e-p'_e$, $q$ being the momentum of the $\gamma^\star$.

\begin{figure}[h]
\label{fkin}
\begin{center}
 \leavevmode
\epsfxsize=0.55\hsize
     \epsfbox{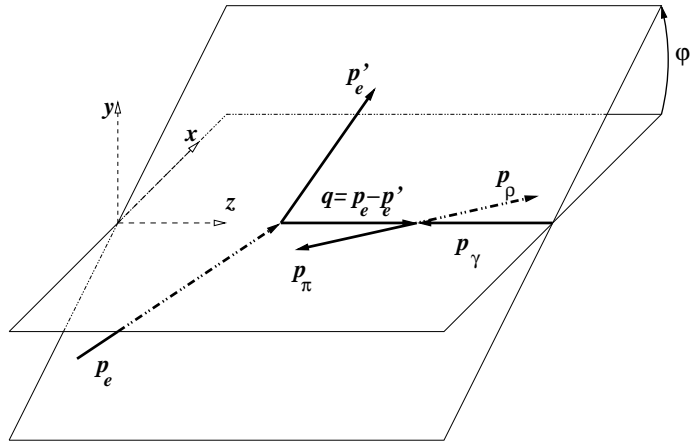}
\end{center}
\caption{ The kinematics of $e(p_e) + \gamma (p_\gamma)
\to e(p'_e) +\rho(p_\rho) + \pi(p_\pi)$ in the center of mass of the meson
pair.}\label{fig:kinematics}
\end{figure}

Then we define the light-cone vectors $p$ and $n$ ($p^2=n^2=0$) 
 such that $2 p.n=1$ and  the invariants in terms of scalar products of the 
momenta:
\begin{center}
\begin{tabular}{ll}
$Q^2=-q^2=-(p_e-p'_e)^2$, &$ \xi= -\frac{\Delta.n}{2P.n}$,\\
$t=(p_{\pi^\mp}-p_\gamma)^2=\Delta^2$,& $s_{e\gamma}=(p_e+p_\gamma)^2$, \\
$W^2=(q+p_\gamma)^2$, & $y=\frac{q.p_\gamma}{p_\gamma.p_e}=\frac{(\xi + 1) Q^2}
{2 \xi s_{e\gamma}}=\frac{Q^2+W^2}{s_{e\gamma} }$. \\
\end{tabular}
\end{center}

\noindent For definiteness, we choose, in the CMS of the meson pair,  
$p=\frac{Q^2+W^2}{2(1+\xi)W}(1,0,0,-1)$ 
and $n=\frac{(1+\xi)W}{2(Q^2+W^2)}(1,0,0,1)$  and we can then express the momenta 
 trough their Sudakov decomposition\footnote{$\Delta_T^2 <0$.}:

\begin{center}
\begin{tabular}{ll}
$p_\gamma= (1+\xi) p$, & $q= \frac{Q^2+W^2}{1+\xi} n - \frac{Q^2}{Q^2+W^2}(1+\xi) p$,\\
$p_{\pi^-}=(1-\xi) p - \frac{\Delta_T^2}{1-\xi} n + \Delta_T$, ~~~~~& 
$ \Delta_T^2=\frac{1-\xi}{1+\xi} t$\\
\end{tabular}
\end{center}

We can see that $\xi$ is 
determined by the external kinematics of $\gamma^\star\gamma \to \rho \pi$
through $\xi\simeq\frac{Q^2}{Q^2 + 2W^2}$ -- similarly to $x_B=\frac{Q^2}{Q^2 + W^2}$ to which it
is linked via the simple relation $\xi\simeq\frac{x_B}{2-x_B}$.

Since we want to focus on the study of the TDA behaviour, we decide to choose 
$Q^2$, $t$, $\xi$ and $\varphi$ as our kinematical variables.  The differential cross section thus reads 
\begin{eqnarray}
\frac{d\sigma^{e \gamma \to e \rho \,\pi}}{dQ^2 dt d\xi d\varphi}= \frac{1}{32 (2\pi)^4 s^2_{e\gamma}}
\frac{1}{\xi (\xi+1)} |{\cal M}^{e \gamma \to e \rho \,\pi}|^2.
\end{eqnarray}

\subsection{  The $ \gamma_L^\star\,\gamma \to   \rho_L^\pm \,\pi^\mp$ amplitude}

Let us first consider the longitudinally polarised $\rho$ meson, 
described by its twist-2 DA. Now the vector character of this DA selects the vector TDA and
the amplitude for  the reaction $\gamma_L^\star \gamma \to \rho_L^+ \,\pi^- $  
becomes  proportional to $V(x, \xi, t)$. It reads
\begin{equation}
\label{amprho} 
{\cal  M}^{TDA}_{\gamma^\star \gamma} (Q^2, \xi,t)=  -\int_{-1}^1 dx \int_{0}^1dz\frac{f_\rho}{f_\pi} \phi_\rho(z)
  M_{h}(z,x,\xi) V(x, \xi, t)\;,
\end{equation}
 where the hard amplitude is ($\bar  z = 1-z$)
\begin{eqnarray}
M_{h}(z,x,\xi) =  
\frac{8\,\pi^2\,\alpha_{em}\,\alpha_s\,C_F}{N_C\,Q}
\frac{1}{z\,\bar z}\left(\frac{Q_u}{x-\xi+i\epsilon} + 
\frac{Q_d}{x+\xi-i\epsilon}  \right)\epsilon^{\mu\nu\rho\sigma}
n_\mu\varepsilon_\nu p_\rho\Delta_\sigma.
\end{eqnarray}   

In the hard subprocess, the quark momenta have been as usual restricted to their component 
collinear w.r.t the associated meson, \ie~for the quark leaving (resp. entering) the 
TDA, $k\simeq (x+\xi) p$
 (resp. $k'\simeq (x-\xi) p$) and the quark (resp. antiquark) entering the $\rho$ meson, 
$\ell\simeq z p_\rho$ (resp. $\ell'\simeq \bar z p_\rho$).

We choose $\phi_\rho(z)=6 z \bar z$   as the asymptotic
normalised meson  distribution amplitude, $f_\rho= 0.216$ GeV~\cite{Ball:1996tb}. After separating 
the real and imaginary part of the amplitude, the $x$-integration gives:
\begin{eqnarray}\label{eq:ivx}
I^V_x&=&\int_{-1}^1 dx \left(\frac{Q_u}{x-\xi+i\epsilon} + 
\frac{Q_d}{x+\xi-i\epsilon}  \right) V(x, \xi, t) \nn\\
&=&Q_u \int^{1}_{-1} dx\,\frac{V(x,\xi,t)-
V(\xi,\xi,t)}{x-\xi}+
Q_d \int^{1}_{-1} dx\, \frac{V(x,\xi,t)-V(-\xi,\xi,t)}{x+\xi}\nn\\
&&+Q_u V(\xi,\xi,t) (\log\left(\frac{1-\xi}{1+\xi}\right)-i\pi)+
Q_d V(-\xi,\xi,t)(\log\left(\frac{1+\xi}{1-\xi}\right)+i\pi ).\nn\\
\end{eqnarray}

The  scaling law for  the amplitude  is 
\begin{equation} 
{\cal M}^{TDA}_{\gamma^\star \gamma}(Q^2, \xi,t) \sim \frac{\alpha_s\sqrt{-t}}{Q}\;,
\label{scaling}
\end{equation} 
up to logarithmic corrections due to the anomalous dimension of the TDA.

\subsection{The $\gamma_L^\star \gamma \to \rho_L^\pm \,\pi^\mp $ contribution to 
$e \gamma \to e \rho_L^\pm \,\pi^\mp $}

The squared amplitude for $e \gamma \to e \rho_L^\pm \,\pi^\mp $ from the subprocess 
$\gamma_L^\star \gamma \to \rho_L^\pm \,\pi^\mp $ is obtained from 
${\cal  M}_{\gamma^\star \gamma}^{TDA} (Q^2, \xi)$
and from the contribution of the fermionic line for the emission of a longitudinal
virtual photon

\beq{}
|{\cal M}_{e\gamma}^{TDA}|^2=\frac{4 \pi \alpha_{em}}{2 Q^4}\tra(\ks p'_e \ks \ep_L(q) \ks p_e \ks \ep^{\star}_L(q)) \;|{\cal M}^{TDA}_{\gamma^\star \gamma}|^2.
\eeq

\begin{figure}[h]
\centering
\includegraphics[width=6cm,angle=-90,clip=true]{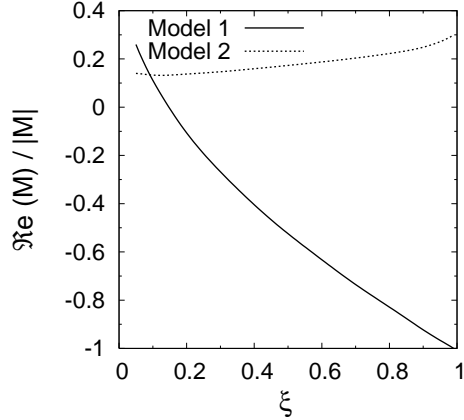}
\caption{$\xi$-dependence of the ratio of the real part to the modulus of the TDA amplitude 
for $t=-0.5$ GeV$^2$ and $Q^2=4$ GeV$^2$. }
\label{fig:realpart}
\end{figure}

Averaging over the real photon polarisation and integrating over $\varphi$ thanks to the 
$\varphi$-independence
of the TDA process, we eventually obtain the differential cross section:
\beq{dsdxidtdq2}
\frac{d\sigma_{e\gamma \to e \rho^+ \pi^-}^{TDA}}{dQ^2 dt  d\xi}=
\frac{128 \pi \alpha_{em}^3 \alpha_s^2  2\pi }
{9 (\xi + 1)^4 Q^8 s_{e\gamma}}\frac{f^2_\rho}{f^2_\pi}(-t)(2 s_{e\gamma} \xi-(\xi + 1) Q^2) (1- \xi) (\re^2(I^V_x)+\im^2(I^V_x)).
\eeq

First, we see that the sole dependence on $t$ is through the factor $-t$ which 
comes from $\epsilon^{\mu\nu\rho\sigma}n_\mu\varepsilon_\nu p_\rho\Delta_\sigma\,$ and physically
accounts for the penalty to turn a helicity $\pm 1$  state into a helicity 0; indeed the cross 
section vanishes at $t=0$. Secondly the positivity of the kinematical factor
$(2 s_{e\gamma} \xi-(\xi + 1) Q^2)$ yields an upper bound for $Q^2$ for given $s_{e\gamma}$ 
and $\xi$. The  $s_{e\gamma}$ dependence of the cross section comes entirely from the factor
$\frac{(2 s_{e\gamma} \xi-(\xi + 1) Q^2)}{s_{e\gamma}}$.

Both real and imaginary part of the amplitude contribute significantly to the cross section, 
which is reasonable at these moderate energies. To quantify this statement, we plot 
on~\cf{fig:realpart} the relative contribution of the real part of the amplitude, 
as a function of the skewness $\xi$. It is independent of the  other variables.

The phenomenological analysis of the pion form factor~\cite{Braun:2005be} seems to indicate that a rather
large ($\sim 1$) value of $\alpha_s$ should be used together with the asymptotic DA. In our plots, 
we therefore put $\alpha_s=1$. Our upcoming conclusions would not be strongly affected by a 
different choice. 

\begin{figure}[h]
\centering
\mbox{\subfigure[]{\includegraphics[width=6cm,angle=-90,clip=true]{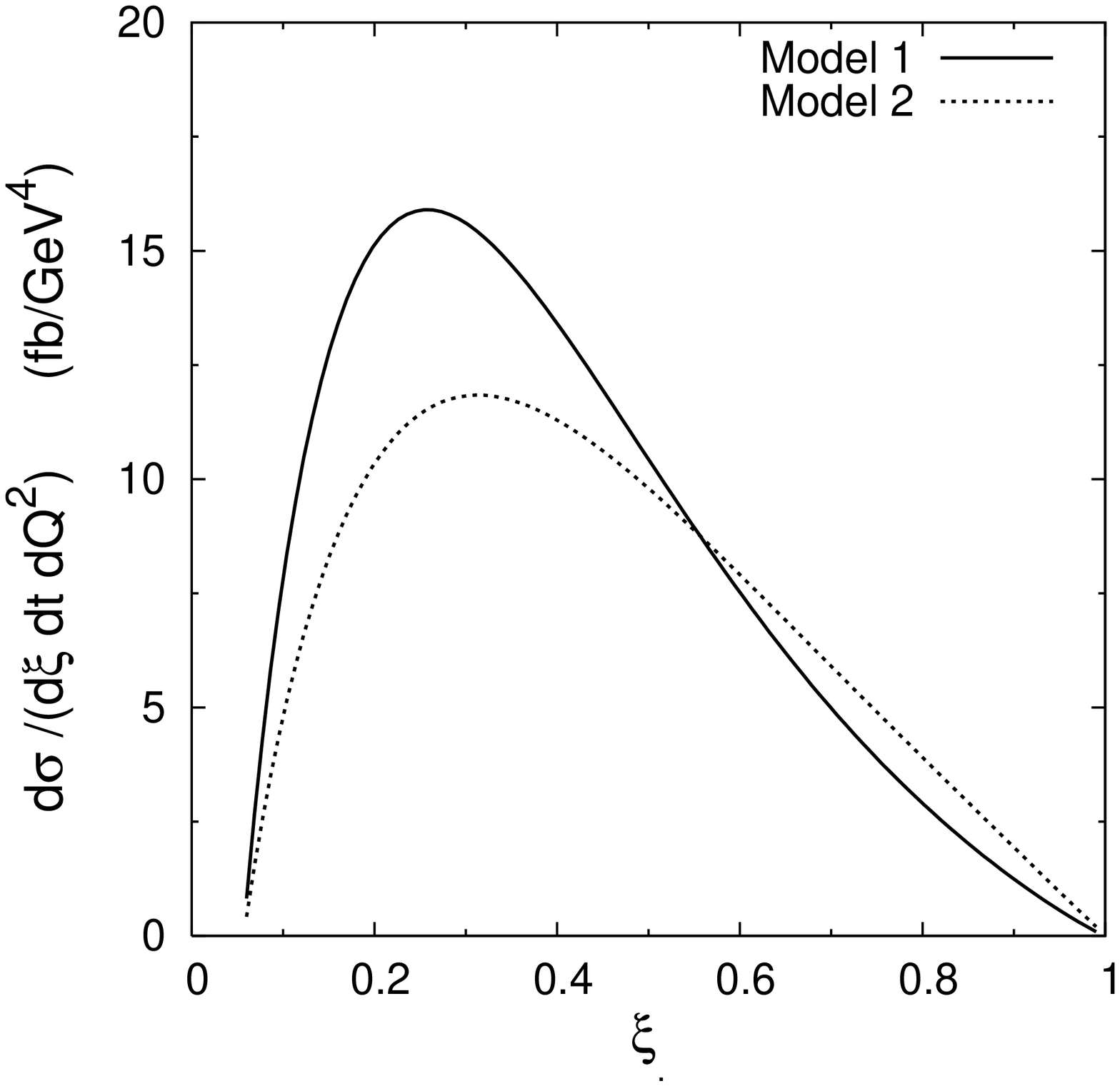}}
      \subfigure[]{\includegraphics[width=6cm,angle=-90]{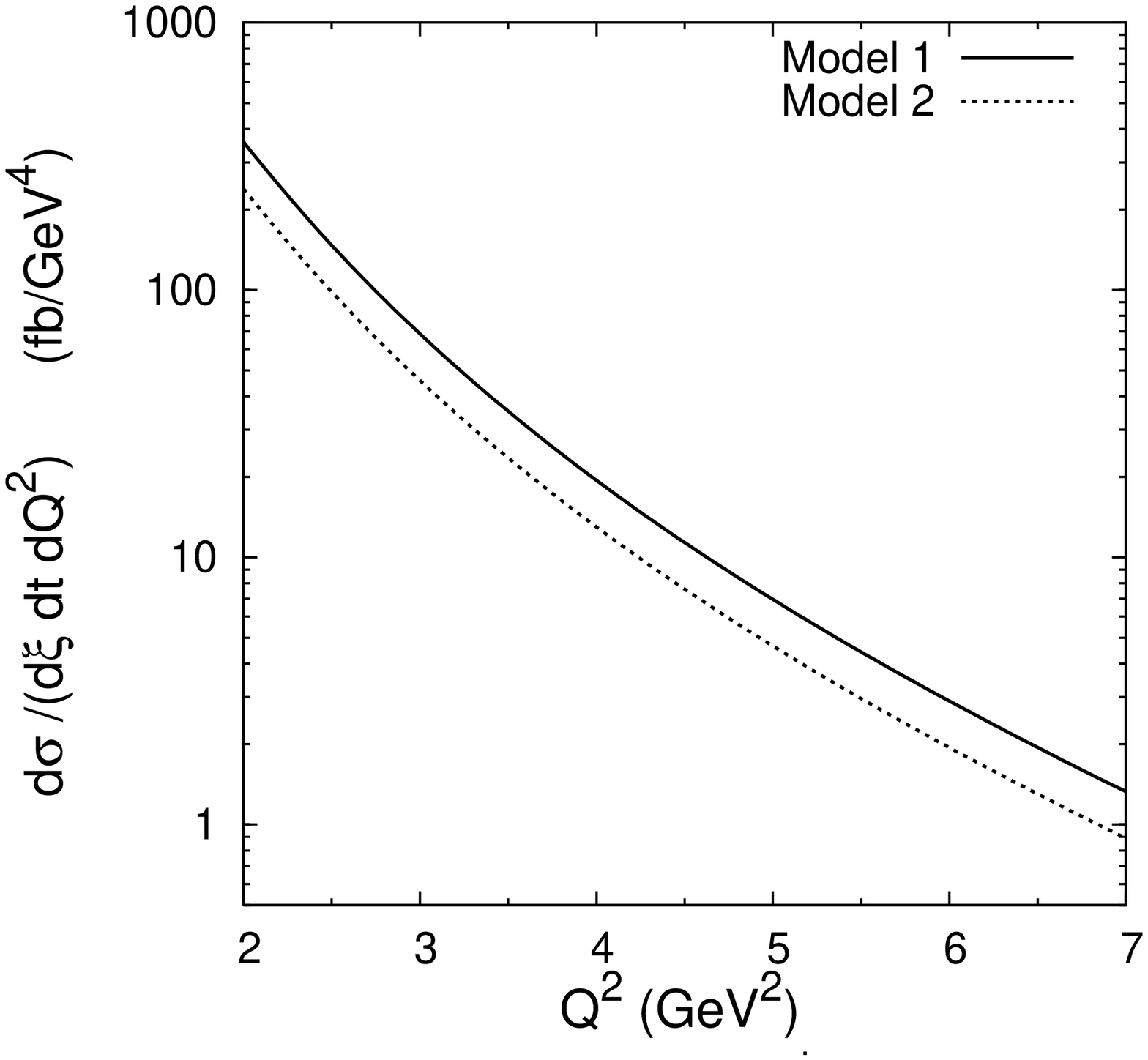}}}
\caption{$e\gamma \to e'\rho^+_L \pi^-$ differential cross section plotted as a function of
$\xi$ (a) and $Q^2$ (b) for $s_{e\gamma}=40$ GeV$^2$, $t=-0.5$ GeV$^2$ and respectively
$Q^2=4$ GeV$^2$ for (a) and $\xi=0.2$ for (b).}
\label{fig:rho_cross_section_xi_q2}
\end{figure}

In \cf{fig:rho_cross_section_xi_q2} (a), we plot the cross section vs $\xi$ and in 
\cf{fig:rho_cross_section_xi_q2} (b) vs $Q^2$ for both models of the vector TDA. 
The $\xi$-dependence  can be generically understood at large
$\xi$ by the factor $(1-\xi)$ and at small values by the limitation on the phase space by the kinematical
factor previously discussed. The behaviour for intermediate values of 
$\xi$ is sensitive to specific models for the TDA. 
As shown in \cf{fig:rho_cross_section_xi_q2} (b), the $Q^2$-behaviour is model independent
and thus constitutes a crucial test of the validity of our approach.

Let us finally add a remark about  the case of transversally polarised $\rho$, 
which could have been quite interesting  since one may naively try to write the amplitude of the 
TDA under the form

\begin{equation}
\label{amp3} 
{\cal  M}_{\gamma^\star \gamma}^{\rho_{T}\pi} (Q^2, \xi) \propto \int dx dz \,\Phi^{\rho_{T}}(z)  M_{h}(z,x,\xi) T(x, \xi, t)\;,
\end{equation}
with the chiral-odd DA for $\rho_T$~\cite{Ball:1996tb} and the tensor chiral-odd TDAs since chiral-odd 
quantities should appear in pairs in a 
physical amplitude. However, a straightforward calculation shows that such an amplitude 
vanishes at leading order due to
the identity $\gamma^\alpha \sigma^{\mu\nu}\gamma_{\alpha} = 0$. This is reminiscent of the 
analysis of transversally-polarised vector-meson electroproduction in the forward region, 
where it has been shown
\cite{CD} that chiral-odd GPD contributions vanish at all orders and the measurement of these GPDs
needs another hard subprocess \cite{IPST}.



\section{ $e \gamma \to e \pi \,\pi $ cross-section  estimates}

\subsection{ $\gamma_L^\star \gamma \to \pi^+ \,\pi^- $ amplitude}

The momenta in the process $e\gamma\to e\pi^+\pi^-$ are defined as in the  
$e \gamma \to e \rho^+ \,\pi^- $ case discussed above,
where the $\pi^+ $ meson flies in 
the direction of the virtual photon 
and the $\pi^-$ meson in the direction of the real one. 

The main difference with the previous case is that the  pion DA replaces the $\rho$ one and that 
the axial  $\gamma \to \pi $ TDA $A(x,\xi,t)$ defined in \ce{eq:def_TDA} replaces the vector one. 
The amplitude thus reads :
\begin{equation}
\label{amppi} 
{\cal  M}_{\gamma^\star \gamma}^{TDA} (Q^2, \xi)=  \int_{-1}^1 dx \int_{0}^1dz \phi_\pi(z)  M_{h}(z,x,\xi) A(x, \xi, t)\;,
\end{equation}
 where the hard amplitude is
\begin{eqnarray}
M_{h}(z,x,\xi) =  
\frac{4\,\pi^2\,\alpha_{em}\,\alpha_s\,C_F}{N_C\,Q}
\frac{1}{z\,\bar z}\left(\frac{Q_u}{x-\xi+i\epsilon} + 
\frac{Q_d}{x+\xi-i\epsilon}  \right)\varepsilon \cdot  \Delta\,,
\end{eqnarray}   
and  where $\phi(z)$ is the normalised meson  distribution amplitude, $\bar  z = 1-z$. The factor $f_\pi$
cancels with the one from the TDA definition.  Now if we choose $\phi^{as}_\pi(z)=6 z \bar z$, 
the $z$-integration is readily carried out and the result has the form of \ce{eq:ivx} 
with obvious replacement of TDAs $V \to A$. The  scaling law for  the amplitude  is 
similar to the one for the $\rho \pi$ case.

\subsection{The $e \gamma \to e \pi^+ \,\pi^- $ cross section}

As said above, the contribution of  the subprocess 
$\gamma_L^\star \gamma \to \pi^+ \,\pi^-$ to the $e \gamma \to e \pi^+ \,\pi^- $ cross section
is obtained from ${\cal  M}_{\gamma^\star \gamma}^{TDA} (Q^2, \xi)$ with the addition
 of the fermionic line for the emission of a longitudinal
virtual photon. The cross section is obtained  from \ce{dsdxidtdq2} by 
the substitutions $I^V \to I^A$ and $f_\rho \to f_\pi$.

The Bremsstrahlung process where the $\pi^+ \,\pi^-$ pair is produced by a photon radiated
from the leptonic line (see the  graphs of~\cf{fig:BH_diagram}) has the following squared amplitude
(omitting for now the interference terms)

\beq{}
\left| {\cal M}_{e\gamma \to e \pi^+ \pi^-}^{\cal B}\right|^2=\frac{64 (4\pi\alpha_{em})^3 \xi^2 }{Q^2 (\xi +1)^2} |F_{\pi}(W^2)|^2 [2+\sqrt{\frac{-t}{Q^2}} A+ \frac{t}{Q^2}B
+\sqrt{\frac{-t}{Q^2}}\frac{t}{Q^2}C+\frac{t^2}{Q^4}D]
\eeq
with
\begin{eqnarray}
A&=& - \cos \varphi    \frac{2   (3 \xi -1) (y-2)}{\sqrt{1+\xi}\sqrt{1-\xi}\sqrt{1-y}}\nn\\
B&=& \frac{[\xi  (5 \xi -2)+1] y^2-2 (y-1) [(1-3 \xi )^2+4 (\xi -1) \xi  \cos 2 \varphi ]}{(\xi -1) (\xi +1) (y-1)}\nn\\
C&=& \cos \varphi  \frac{4 \xi (3 \xi -1) (y-2)}{ (1+\xi)\sqrt{1+\xi}\sqrt{1-\xi}\sqrt{1-y} }\nn\\
D&=&\frac{8\xi^2}{(\xi +1)^2}\nn
\end{eqnarray}

\begin{figure}[h]
\centering
\mbox{\subfigure[direct]{\includegraphics[width=5cm]{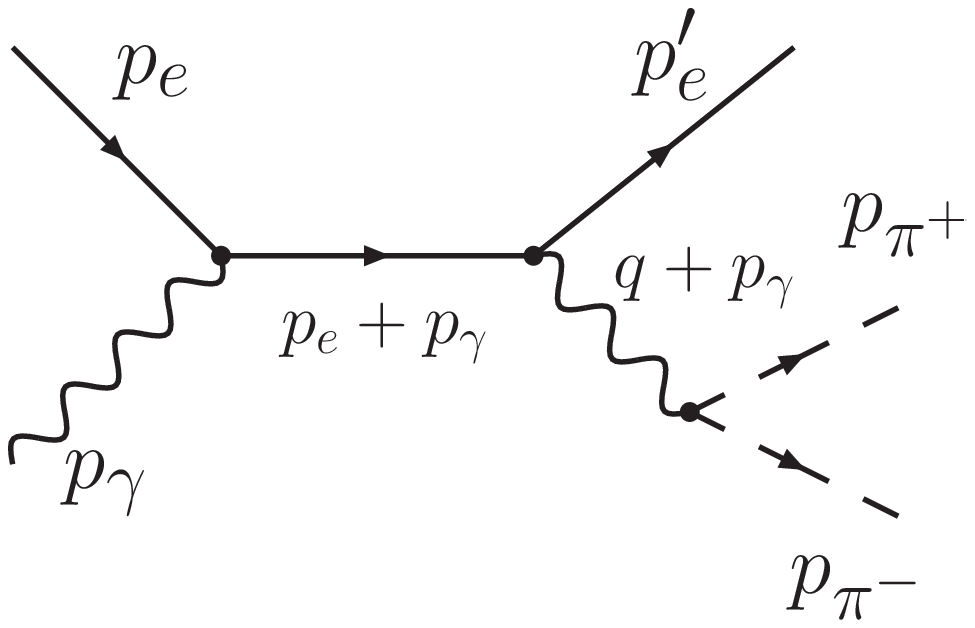}}\quad\quad
      \subfigure[crossed]{\includegraphics[width=5cm]{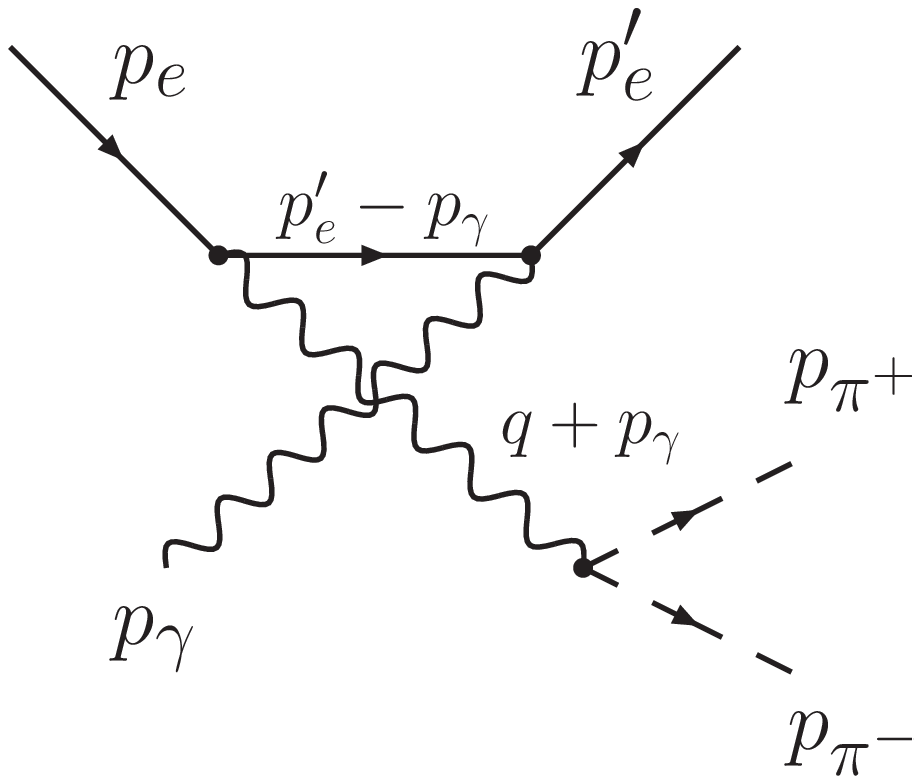}}}
\caption{Feynman graphs for the Bremsstrahlung contribution to $e \gamma \to e \pi^+ \,\pi^- $ at LO.}
\label{fig:BH_diagram}
\end{figure}

The factor
used at the $\pi^+\pi^-$ vertex is $i e (p_{\pi^-}-p_{\pi^+})^\mu F_{\pi}((q+p_\gamma)^2)$ where
$F_{\pi}(W^2)$ is the pion form factor which is well measured for $W^2$ up to a few GeV$^2$
and is described in a satisfactorily way at large space-like values by perturbative QCD~\cite{ERBL}, 
while the time-like region is less understood~\cite{GP} but is measured~\cite{Pedlar:2005sj}.

In the following estimates, and to be consistent with our previous choice of the asymptotic pion DA 
in the TDA subprocess amplitude, we describe the pion form factor $F_{\pi}(W^2)$ with the asymptotic 
distribution amplitude and a large value of $\alpha_s$. This choice will not affect the main 
conclusions of this section, and the inclusion of a fit to the available data would give the 
same order of magnitude for the Bremsstrahlung contribution, which will remain mostly negligible 
in all the kinematical domain considered. 

The use of the asymptotic form for the pion DA has been criticised and other inputs have been 
proposed on the basis of QCD sum rules~\cite{Chernyak:1983ej}. If one uses the C-Z parametrisation and
choose $\alpha_s=0.6$ , the amplitude remains unchanged. This has to be paralleled with the  form factor 
analysis, where the use of  the C-Z DA is sometimes accompanied 
by the use of a lower value for the strong coupling constant $\alpha_s$ of the order of 
0.4~\cite{Braun:2005be}.

 We show in~\cf{fig:ds_pi} the relative contributions of the Bremsstrahlung and TDA subprocesses 
(omitting interference terms) to the differential cross section integrated over $\varphi$,
 as a function of $s_{e\gamma}$, $\xi$ and $Q^2$. Except for small values 
of $s_{e\gamma}$, the TDA subprocess clearly dominates for all reasonable values of $\xi$. 
We also show the $\varphi$-dependence of the Bremsstrahlung contribution, which may be turned 
into a positive check of the dominance of the $\varphi$ independent TDA subprocess, 
much in the manner of the successful test of the GPD framework in deeply virtual Compton 
scattering~\cite{DGPR}.

Since the Bremsstrahlung contribution is quite small, we have not given the 
expression of the interference terms which may be anyhow cancelled out if one 
considers the charge-averaged cross section
$d\sigma (\pi^+ \pi^-) +d\sigma (\pi^- \pi^+)$ since the TDA and Bremsstrahlung amplitudes 
produce hadronic states with opposite charge conjugations. On the other hand,  this latter 
property may  be used to separate the interference contribution by measuring the charge 
asymmetric quantity $d\sigma (\pi^+ \pi^-) - d\sigma (\pi^- \pi^+)$ , and thus analysing 
the TDA subprocess at the amplitude level~\cite{HPST}.

\begin{figure}[H]
\centering
\mbox{\subfigure{\includegraphics[width=6cm,angle=-90]{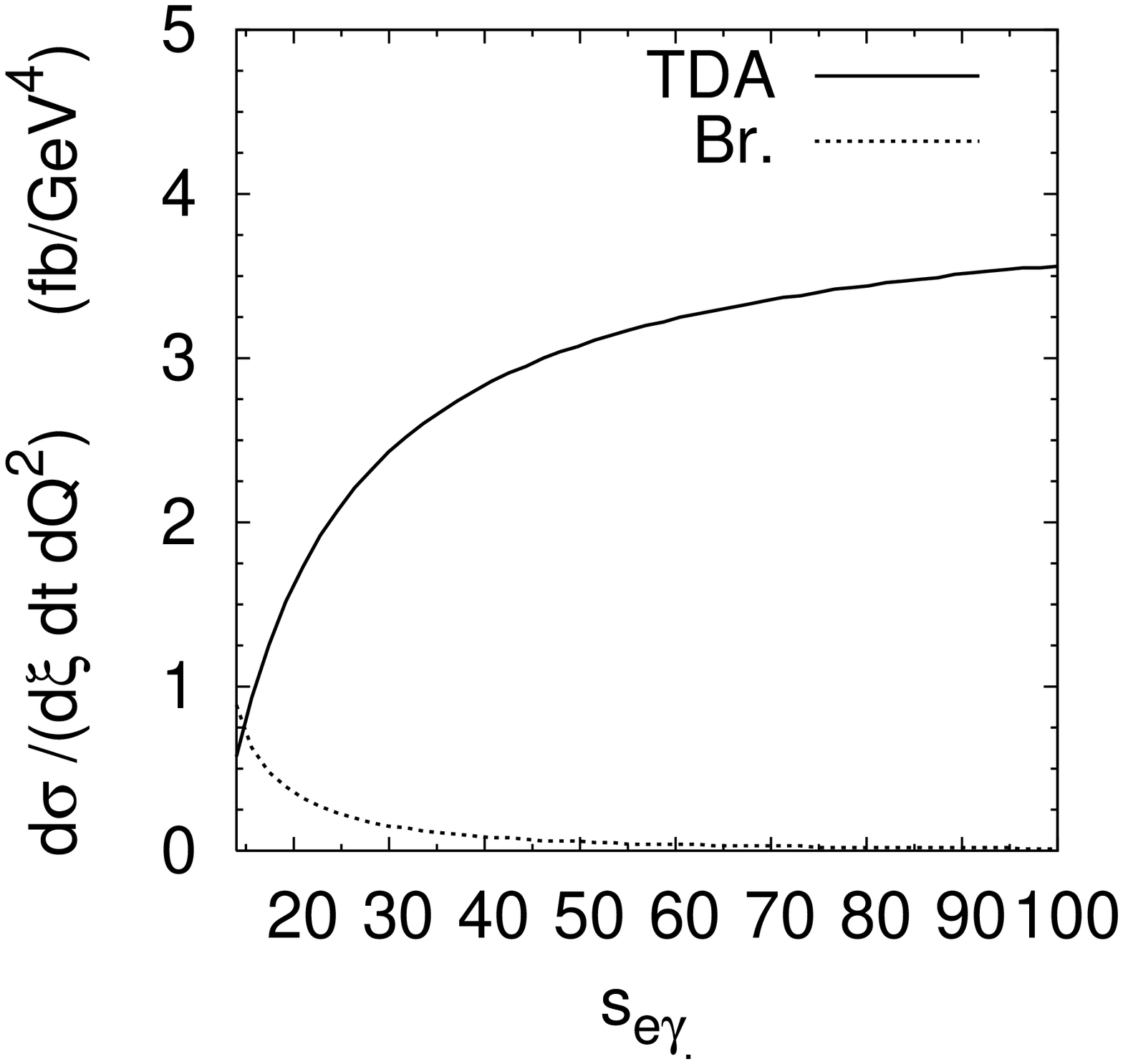}}
      \subfigure{\includegraphics[width=6cm,angle=-90]{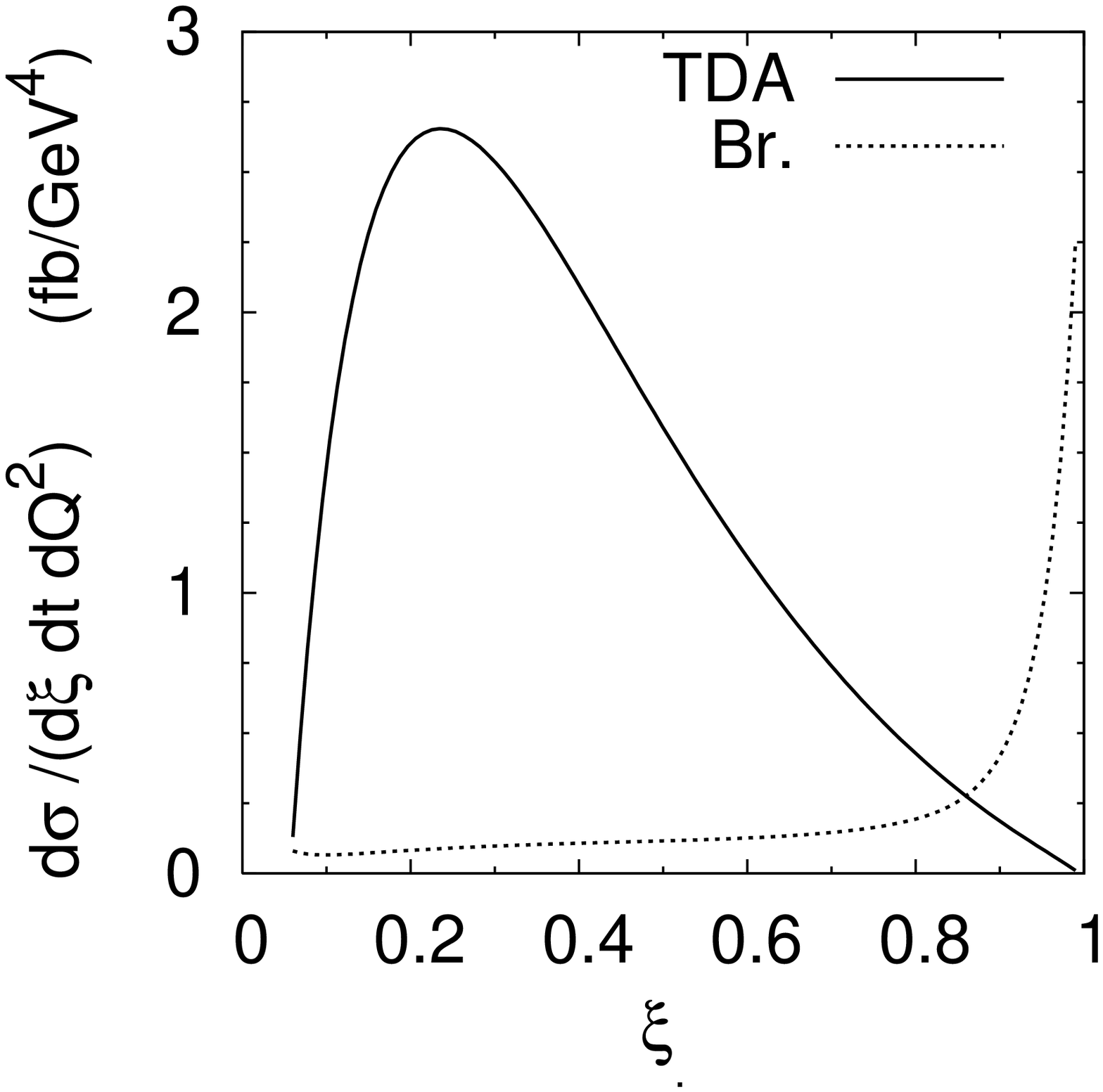}}}
\mbox{ \subfigure{\includegraphics[width=6cm,angle=-90]{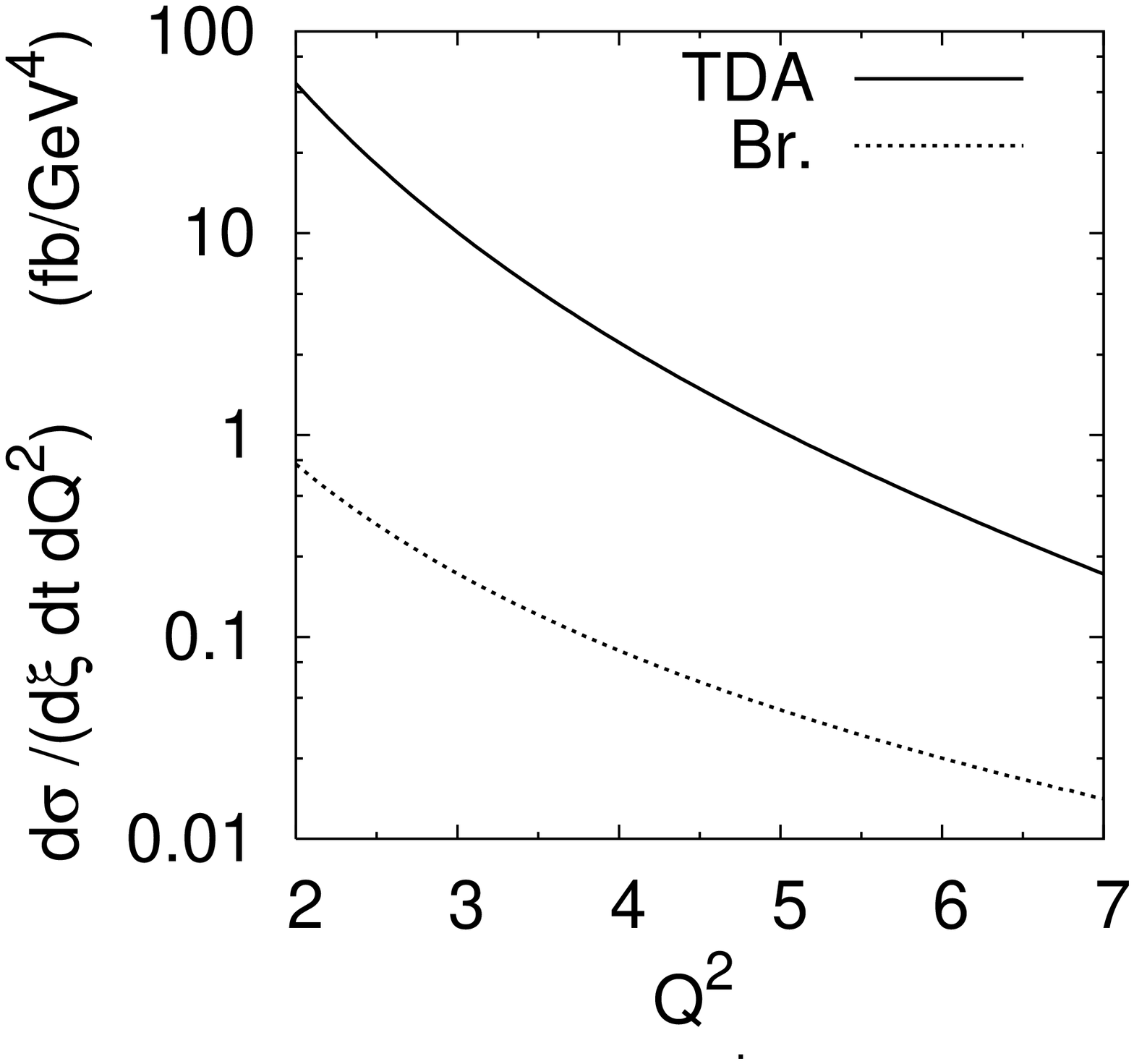}}
\subfigure{\includegraphics[width=6cm,angle=-90]{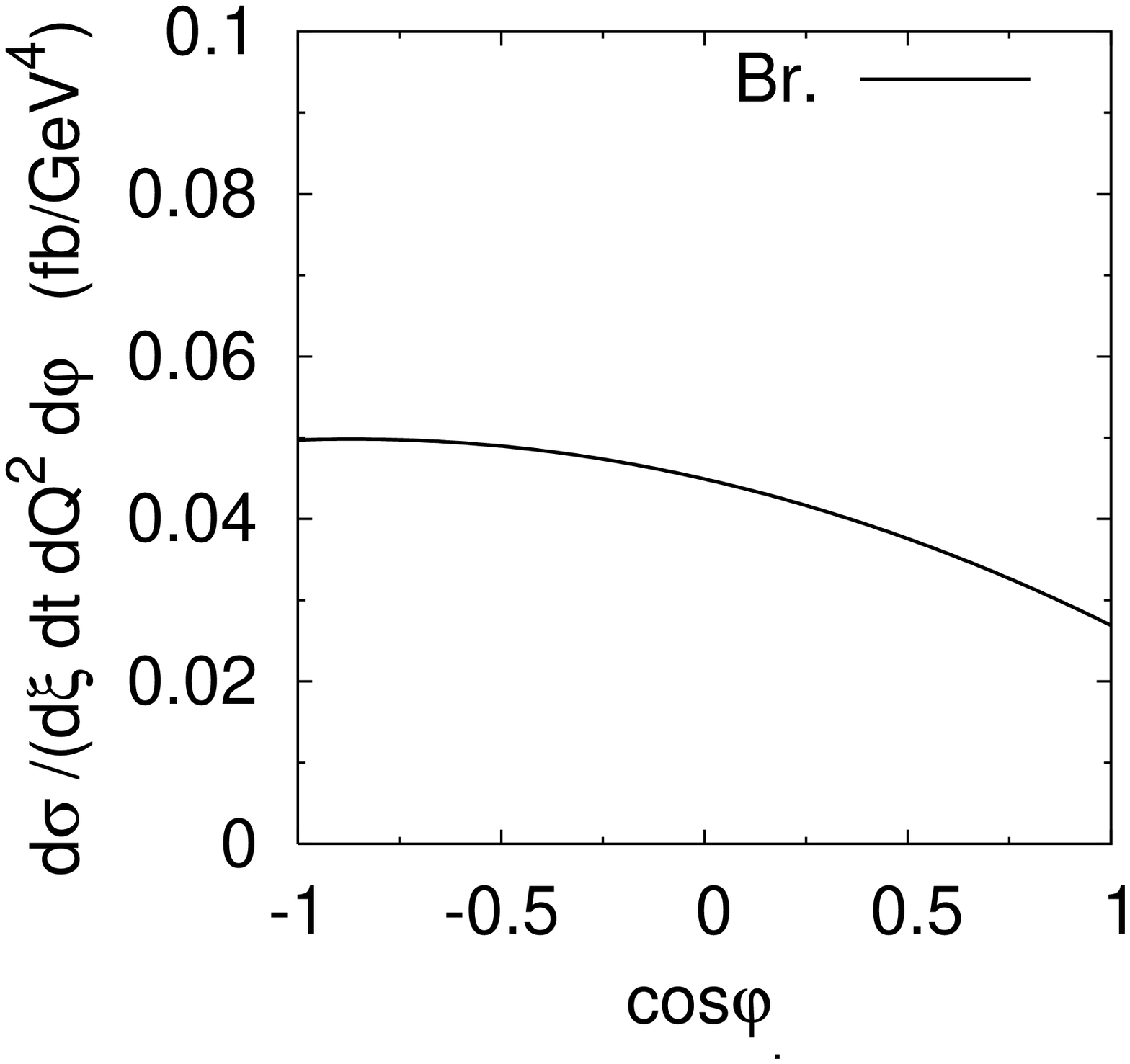}}
}
\caption{Differential cross sections for the TDA and the Bremsstrahlung subprocesses
as a function of $s_{e\gamma}$, $\xi$ and $Q^2$, and for the sole Bremsstrahlung subprocess
as a function of $\cos \varphi$ (lower right plot). Except for the variable studied, 
$Q^2=4$ GeV$^2$, $\xi=0.2$, $t=-0.5$ GeV$^2$ and  $s_{e\gamma}=40$  GeV$^2$.}
\label{fig:ds_pi}
\end{figure}

\subsection{ The neutral case: $e \gamma \to e \pi^0 \,\pi^0 $}

In this case, $C$ invariance forbids any contribution from the Bremsstrahlung subprocess.
We emphasise here that the measurement of the cross section for $e \gamma \to e \pi^0 \,\pi^0 $
in the forward region would provide one of the sole possible determination of
the axial form factor $F_A$ of the $\pi^0$ at small $t$. Indeed, its measurement is not feasible in
$ \pi^0 \to  \nu \bar \nu \gamma  $ and it is completely drown in the electromagnetic
background in $ \pi^0 \to  e^+ e^- \gamma$. 

Therefore, any measurement of the cross section for $e \gamma \to e \pi^0 \,\pi^0 $
would constrain the value of $F_A^{\pi^0}$ via the sum rule for the axial
TDAs $A^{\pi^0}_q(x,\xi,t)$ similar to the one linking $F_{\pi^0\gamma^\star \gamma}$ to 
$V^{\pi^0}_q(x,\xi,t)$ (see~\ce{srVpi0}). We expect the cross section
$\gamma^\star \gamma \to \pi^0 \,\pi^0 $ to be of the order of  the one for 
$\gamma^\star \gamma \to \pi^\pm \,\pi^\mp $ depicted in~\cf{fig:ds_pi}.


\section {Conclusion}

In summary, we have shown how to test   experimentally  the new factorised QCD approach to forward
 hard exclusive scattering. We believe that our models for the photon to meson transition 
distribution amplitudes are sufficiently constrained to give reasonable orders of magnitude for
 the estimated cross sections. The good news is that the hard hadronic process dominates the 
Bremsstrahlung contribution in the kinematics which are accessible in existing $e^+ e^-$ colliders. 
The goal is now to test our approach, in particular by verifying the scaling of the cross sections, 
and then to measure these new hadronic matrix elements. Cross sections are large enough for quantitative 
studies to be performed. On the other hand, lattice studies could calculate these
TDAs, at least within some approximations. Chiral perturbation theory may even be used to improve
the extrapolation from a large quark mass to a realistic model. Higher-order corrections to the 
hard scattering process should be studied. The long history of the improvements of the QCD 
understanding of form factors at large momentum transfers~\cite{thff} should give us a way 
to better include 
the effects of \eg~end point regions and their partial Sudakov suppressions. 
As in all studies of this type, possible higher-twist contributions may be relevant
at measurable values of $Q^2$. Some studies have been done in related cases in particular models~\cite{HT}.
Experimental verification of the scaling laws are the first tests to be applied to our description. 
Moreover, the kinematical domain of applicability of the present description (and in particular the 
small transfer, large energy, requirements) may not be clear cut in real experiments using existing 
accelerators. An interesting further study 
should be to try to understand the transition regions between the kinematics where 
our framework should apply 
on the one hand and on the other hand the other existing QCD descriptions 
 of the same $\gamma^\star \gamma \to  A B $  reactions at 
small energy~\cite{GDAAPT} or at very large energy~\cite{PSW}.

\subsection*{Acknowledgments.}

We are thankful to M.~Diehl, B.~Moussallam, T.N.~Pham, O.~P\`ene, C.~Roiesnel, 
J.~Stern and S.~Wallon for useful discussions and correspondence.
This work is partly supported by the French-Polish scientific agreement Polonium, 
the Polish Grant 1 P03B 028 28, the
Fonds National de la Recherche Scientifique (FNRS, Belgium), the ECO-NET program, contract 
12584QK and the Joint Research Activity "Generalised Parton Distributions" of the european I3 program
Hadronic Physics, contract RII3-CT-2004-506078.

\small


\end{document}